\documentclass[12pt]{iopart}

\usepackage{graphicx}
\usepackage{dsfont}
\usepackage{iopams}
\usepackage{dsfont}
\usepackage{epstopdf}
\usepackage{subfig}

\newcommand{\ket}[1]{\left|#1\right>}

\begin{document}

\title{Double self-Kerr scheme for optical Schr\"odinger-cat state preparation}

\author{P. Adam$^{1,3}$, Z. Dar\'azs$^{1,2}$, T. Kiss$^{1}$ and M. Mechler$^{3}$}
\address{
$^1$Research Institute for Solid State Physics and Optics, HAS, Konkoly-Thege M. u. 29-33, H-1121 Budapest, Hungary \\
$^2$E\"otv\"os University, P\'azm\'any P\'eter s\'et\'any 1/A, H-1117 Budapest, Hungary\\
$^{3}$Department of Physics, University of P\'ecs, Ifj\'us\'ag u. 6., 7624 P\'ecs, Hungary
}

%\pacs{03.67.-a,05.40.Fb,02.30.Mv}

\ead{adam@szfki.hu}

\begin{abstract}
We propose a scheme to prepare optical Schr\"odinger-cat states in a traveling wave setting. Two states are similarly prepared via the self-Kerr effect and after mixing them, one mode is measured by homodyne detection. In the other mode a superposition of coherent states is conditionally prepared. The advantage of the scheme is that assuming a small Kerr effect one can prepare with high probability one from a set of Schr\"odinger-cat states. The measured value of the quadrature provides the information, which state from the set is actually prepared.
\end{abstract}

\section{Introduction}
Superpositions of coherent states (CSS) are often referred to as Schr\"odinger-cat states, as the constituents of the superposition are two quasi-classical states. It would be desirable to prepare quantum systems in such states both for practical and fundamental purposes. Their non-classical properties could be exploited e.g. in quantum communication. 
In traveling wave optics there are a number of proposals to prepare Schr\"odingar-cat states \cite{Glancy}.

Photon subtraction from a squeezed vacuum state is a successfully applied method to produce approximate CSS states \cite{Dakna,CSSexperiments}. Recently, squeezed CSSs were also generated with similar technique \cite{CSSexperiments2}. In another proposal for conditional preparation, squeezed single photon states are let interfere on a beamsplitter and states in one arm are conditionally selected by photon detection \cite{Lund2004}.

Propagation through nonlinear media exhibiting the Kerr effect was suggested in one of the first proposals to prepare CSS states \cite{Yurke}. There has been several proposals employing the cross-Kerr effect \cite{CrossKerr}, as well as the self-Kerr effect \cite{SelfKerr}. A double cross-Kerr scheme have recently been proposed by us \cite{Scripta}, exploiting the interference of two similar states on a beam splitter for a conditional CSS preparation scheme.

In this paper, we propose a scheme with two identically prepared states by propagating a traveling wave coherent state through a self-Kerr medium. The states are let interfere on a balanced beam splitter then one of them measured by balanced homodyne detection.

\section{Double self-Kerr scheme}

The proposed scheme is depicted in Figure 1. Two identical states are prepared from identical coherent states by letting them propagate through a medium exhibiting self-Kerr effect. 
\begin{figure}[ht!]
\label{fig:scheme}
\center
\resizebox{0.77 \linewidth}{!}{
 \includegraphics{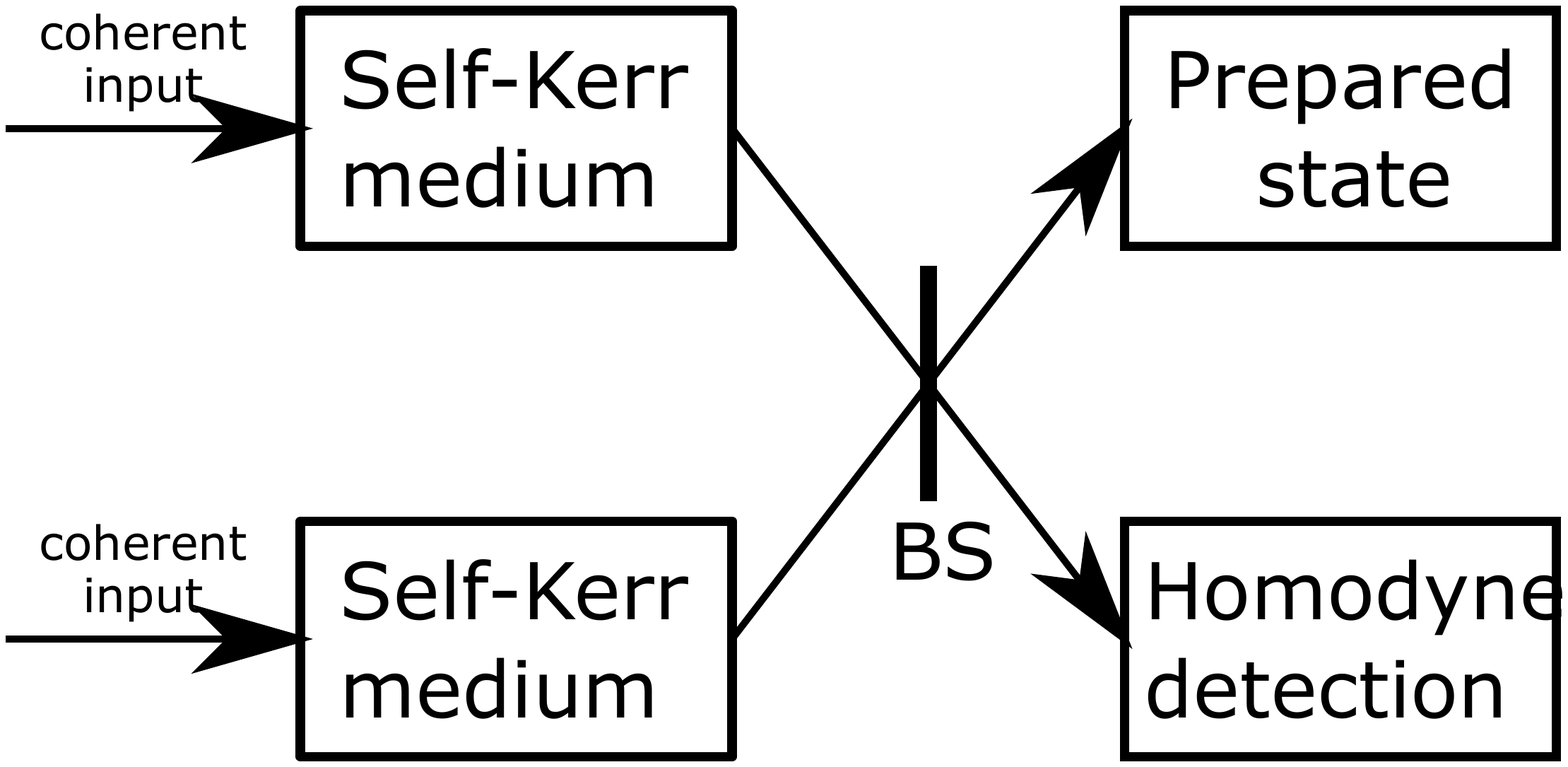}
}
 \caption{The proposed scheme. Two coherent states with the same amplitude are fed into media exhibiting self-Kerr effect. The two identically prepared states are then mixed on a balanced beam splitter. One of the outputs is measured by balanced homodyne detection.}
\end{figure}
After the interaction both modes are prepared in the state \cite{KerrState}
\begin{equation}
|\Psi\rangle= \sum_{k=1}^N c_k |\alpha_{k}\rangle \, , \quad \alpha_{k}=\alpha i\exp( i 2 k \pi/N )\, ,
\end{equation}
where 
\begin{equation}
c_k=\frac{e^{i\xi_k}}{\sqrt{N}}=\frac{1}{N} \sum_{l=0}^{N-1}(-1)^l \exp \left [   -\frac{i\pi l}{N} (2k-l)  \right ] \, .
\end{equation}
The balanced beam splitter transforms coherent states in the following way:
\begin{equation}
|\alpha\rangle_1 |\beta\rangle_2 \to \left |   \frac{\alpha+\beta}{\sqrt{2}}   \right \rangle_3   \left | \frac{\alpha-\beta}{\sqrt{2}}   \right \rangle_4 \, .
\end{equation}
The two-mode state at the outputs of the beam splitter reads
\begin{equation}
\label{eq:two-mode}
|\Phi \rangle _{3,4}=\sum_{k=1}^N \sum_{l=1}^N c_kc_l \left | \frac{\alpha_k +\alpha_l}{\sqrt{2}} \right \rangle_3\left | \frac{\alpha_k -\alpha_l}{\sqrt{2}} \right \rangle_4 \, .
\end{equation}
In mode 3 we measure by balanced optical homodyne detection the real quadrature. 
If the measurement results in the value $|X_{m}\rangle$, then the state after the homodyne detection in mode 4 will have the form
\begin{equation}
|\phi\rangle = \mathcal{N}_{\phi }\sum_{k=1}^N \sum_{l=1}^N c_kc_l \left \langle X_{m} \left | \frac{\alpha_k +\alpha_l}{\sqrt{2}} \right \rangle \right .  dX\left | \frac{\alpha_k -\alpha_l}{\sqrt{2}} \right \rangle \, .
\end{equation}
The resulting state can be written as a superposition of CSS states and the vacuum
\begin{eqnarray}
\label{eq:detected}
|\phi  \rangle&=&
\mathcal{N}_{\phi  } \left [  \left ( \sum_{k=1}^N  c_k^2
\langle X_{m}|\sqrt{2}\alpha_k\rangle dX \right ) |0\rangle + \right . \nonumber \\
& &  \left . \sum_{k,l=1, \, k<l}^N  \frac{c_kc_l }{\mathcal{N}_{CSS}} \left \langle X_{m} \left | \frac{\alpha_k+\alpha_l}{\sqrt{2}} \right \rangle \right . dX |\textrm{CSS}_{k,l}\rangle \right ] \, ,
\end{eqnarray}
where we have defined the CSS states
\begin{equation}
|\textrm{CSS}_{k,l}\rangle=\mathcal{N}_{CSS}\left ( \left | \frac{\alpha_k -\alpha_l}{\sqrt{2}} \right \rangle +  \left | \frac{\alpha_l -\alpha_k}{\sqrt{2}} \right \rangle \right ) \, ,
\end{equation}
and the normalization factor reads
$\mathcal{N}_{CSS}=(2+2 \exp (-|\alpha_{k}-\alpha_{l}|^2))^{-1/2}$.
The CSS states may be distinguished only if we choose $N=2n+1$. Further conditions of distinguishability stem from the requirement that the overlaps of the Gaussian quadrature wave functions in mode 3 are minimal.

The constituent coherent states of the resulting CSS states in mode 4 form $n$ circles in phase space. This can be easily seen by calculating the 
absolute values of the coherent state amplitudes.
The radii of the circles read
\begin{equation}
\frac{|\alpha_k-\alpha_l|}{\sqrt{2}}=\alpha\sqrt{1-\cos \frac{2(k-l)\pi}{2n+1}} \, .
\end{equation}
The largest circle has a radius smaller, but close to $\sqrt{2}\alpha$.

\section{An example: $N=5$}

\begin{figure}[ht!]
 \begin{center}
\subfloat{\includegraphics[width=0.45\textwidth ]{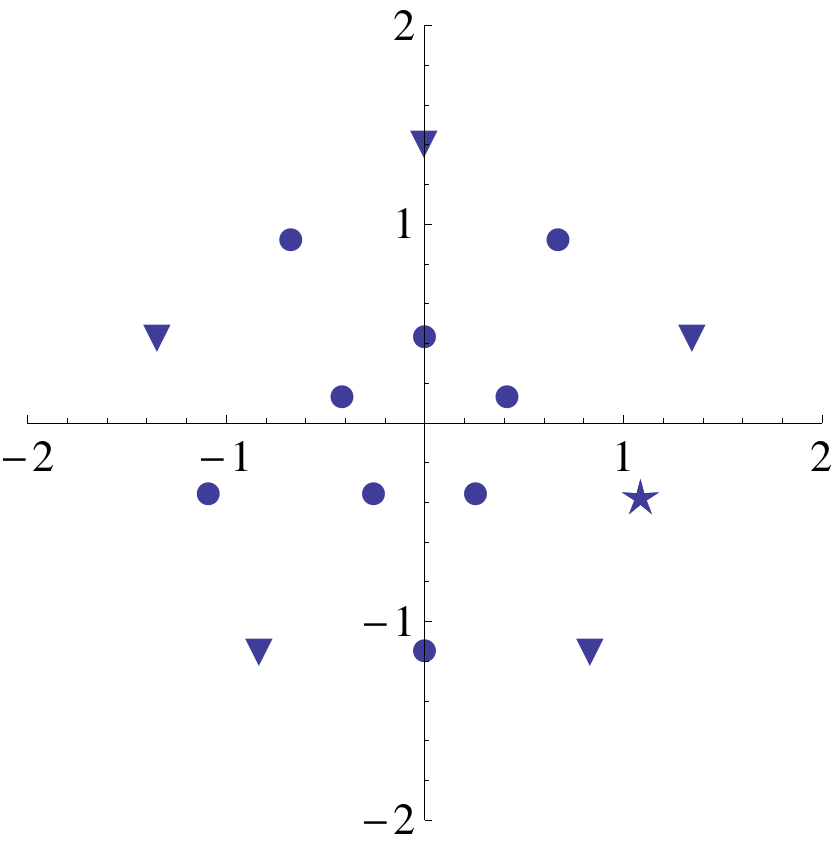}}
\subfloat{\qquad}
\subfloat{\includegraphics[width=0.45\textwidth ]{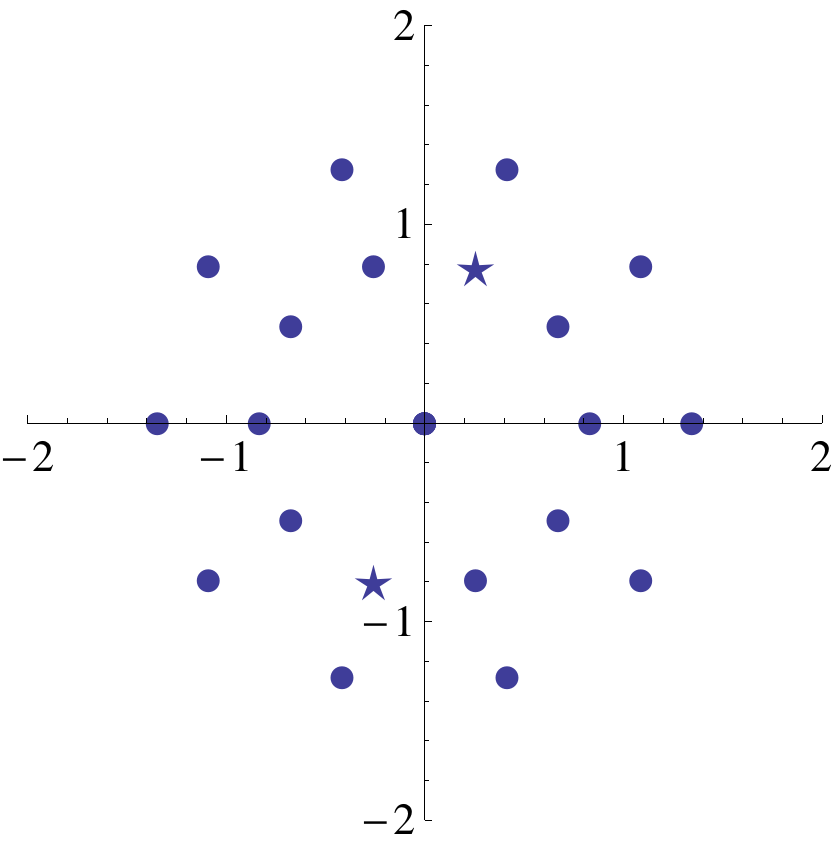}}
\end{center}
\caption{Illustration of the amplitudes of the coherent state in both modes 3 and 4. Filled circles state on the left (mode 3) are  multiplied by CSS states on the right (mode 4). The star denotes the target state to be detected in mode 3 (left) and the corresponding prepared CSS state (right).}
\label{fig:five}
\end{figure}

\begin{figure}[ht!]
\center
\resizebox{100mm}{!}{\includegraphics{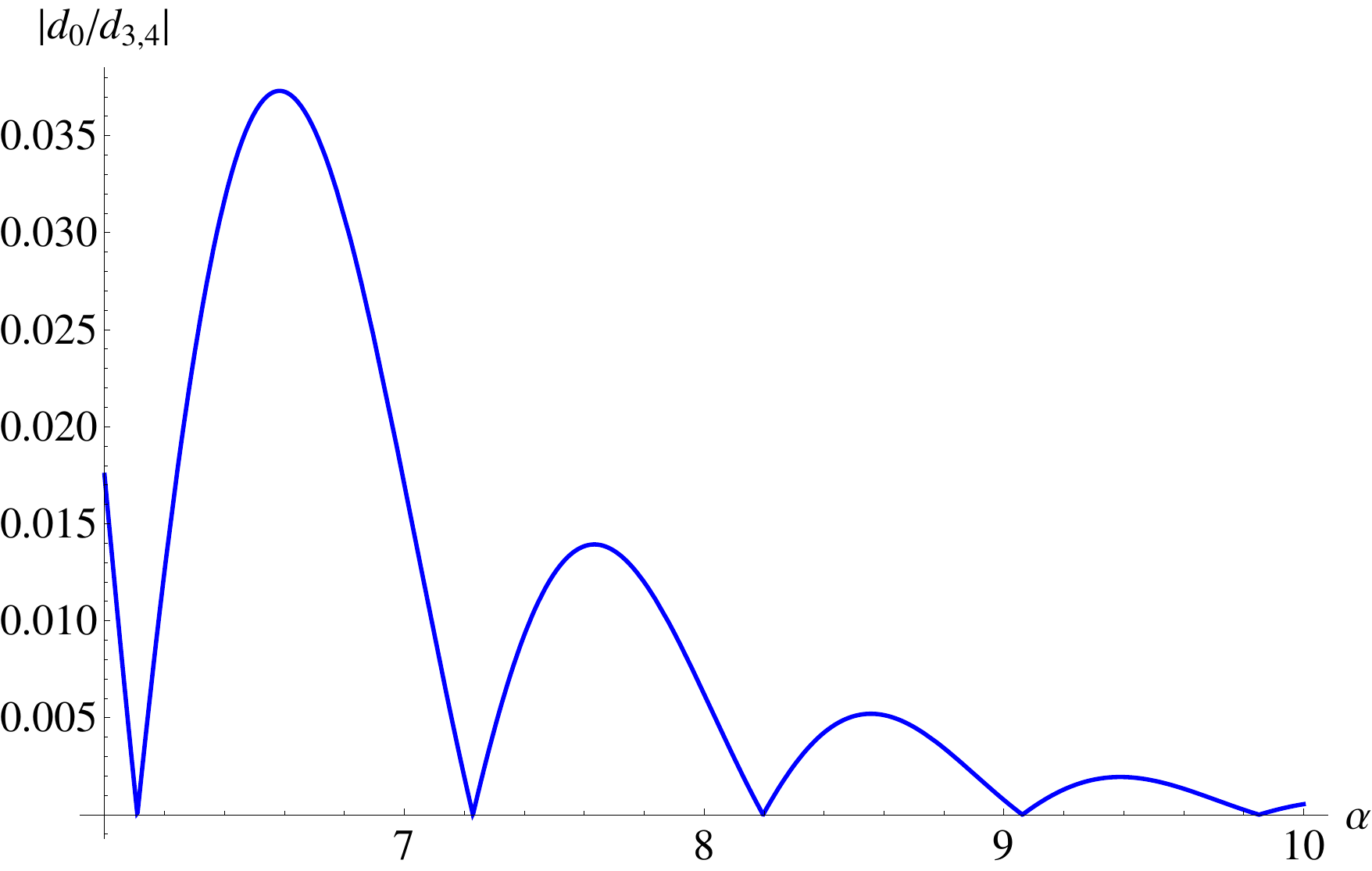}}
\caption{The relative amplitude of the vacuum and the desired CSS state is shown in dependence of the original coherent state amplitude. Due to destructive interference, at certain values of the amplitude $\alpha$ the extra vacuum term will be completely missing from the prepared superposition.
}
\label{fig:alpha}
\end{figure}

In this section we demonstrate on a concrete example how the proposed scheme works. We choose a moderate number $N=5$, in order to explicitly show the working mechanism of selection. In practice, a weak self-Kerr effect might require to work with higher values of $N$, but the principle for preparation will be the same.
The state incident on both inputs of the beam splitter reads
\begin{equation}
\ket{\Phi_5 }= c_1\ket{\alpha e^{\frac{9 i \pi }{10}}} + c_2\ket{\alpha e^{-\frac{7 i \pi }{10}}} + c_3\ket{\alpha e^{-\frac{3 i \pi }{10}}} + c_4\ket{\alpha e^{\frac{i \pi }{10}}} +    c_5\ket{\alpha i} \, .
\end{equation}
The beam splitter transforms the states according to Eq. \ref{eq:two-mode}.
We illustrate the complex amplitudes of the participating coherent states for both modes in Fig. (\ref{fig:five}).
The state after homodyne detection will be a superposition of the vacuum and 10 different CSS states
\begin{equation}
|\phi_5 \rangle=
\mathcal{N}_{\phi  } \left [ d_{0} |0\rangle 
+   
\sum_{k,l=1, \, k<l}^5 d_{kl}|\textrm{CSS}_{k,l}\rangle \right ] \, .
\end{equation}
The coefficients $d$ can be determined from Eq. \ref{eq:detected}.
The CSS states in mode 4, after the homodyne measurement of the value $X_{m}$ in mode 3, are multiplied by the constants 
\begin{equation}
d_{k,l}=\frac{c_kc_l }{\mathcal{N}_{CSS}}  \left \langle X_m \left | \frac{\alpha_k+\alpha_l}{\sqrt{2}} \right . \right \rangle dX .
\end{equation}
The actual ratio of the constants depend on the value of the measured quadrature. A good choice for homodyne measurement is to pick the most isolated coherent state. For $N=5$ the best choice is to measure around the peak corresponding to the coherent state with amplitude
\begin{equation}
\alpha_{3,4}=\frac{\alpha_3+\alpha_4}{\sqrt{2} } =
\frac{e^{\frac{i \pi }{10}}+e^{-\frac{3 i \pi }{10}}}{\sqrt{2}} \, .
\end{equation}
In this way the state $|\textrm{CSS}_{3,4}\rangle$ will be approximately prepared.
In mode 4 the constant multiplying the vacuum reads 
\begin{equation}
d_0= \sum_{k=1}^N  c_k^2 \langle X_{m}|\sqrt{2}\alpha_k\rangle dX   \, .
\end{equation}
Examining its absolute value as a function of $\alpha$, we find that it exhibits oscillations, and at certain points it can be exactly zero. The function is depicted in Fig. \ref{fig:alpha}.

Let us fix, as an example, $\alpha=7.23$. With this choice the amplitudes of the other states in the superposition are negligible compared to the target CSS state (the largest ones have the following numerical values: $|d_{0}/d_{3,4}|=9.0 \cdot 10^{-4}$, $|d_{4,5}/d_{3,4}|=1.2 \cdot 10^{-4}$). We note that with increasing the amplitude $\alpha$
the weight of the other CSS state in the superposition decays monotonically, thus by the appropriate choice of the intensity of the input coherent state, one can achieve high fidelity preparation.

\section{Conclusions}

We have proposed a conditional scheme to prepare coherent state superpositions. The scheme is based on a double self-Kerr scheme, where two identically prepared states interfere on a balanced beam splitter and subsequently one of the outputs is measured by balanced homodyne detection. The detection event in general prepares a superposition of CSS states and the vacuum. We have shown that due to destructive interference the vacuum amplitude can be zero, depending on the amplitude 
of the initial coherent state and the selected CSS state. 
The separation of the selected CSS state can vary, conditioned on the measurement result.

\ack
The financial support by the Czech-Hungarian cooperation project (CZ-11/2009) is gratefully acknowledged.


\section*{References}

\begin{thebibliography}{x}

\bibitem{Glancy}
Glancy S and Vasconcelos H M 2008 \JOSA B \textbf{25} 721

\bibitem{Dakna}
Dakna M, Anhut T, Opatrn\'y T, Kn\"oll L and Welsch D-G 1997 \PR A \textbf{55} 3184

\bibitem{CSSexperiments}
Neegaard-Nielsen J S, Melholt Nielsen B, Hettich C, Molmer K and Polzik E S 2006 \PRL \textbf{97} 083604

\bibitem{CSSexperiments2}
Ourjoumtsev A, Jeong H, Tualle-Brouri R and Grangier P 2007 Nature \textbf{448} 784;
Takahashi H, Wauki K, Suzuki S, Takeoka M, Hayasaka K, Furusawa A and Sasaki M 2008 \PRL \textbf{101} 233605

\bibitem{Lund2004}
Lund A P, Jeong H, Ralph T C and Kim M S 2004 \PR A \textbf{70} 020101

\bibitem{Yurke} Yurke B,	 Stoler D 1986 \PRL \textbf{57} 13;
Yurke B Stoler D 1988 \textit{Physica B+C} \textbf{151} 298

\bibitem{CrossKerr} Gerry C C 1999 \PR A \textbf{59} 4095;
Jeong H 2005 \PR A \textbf{72} 034305;
He B, Nadeem M and Bergou J A 2009 \PR A \textbf{79} 035802

\bibitem{SelfKerr}
Jeong H, Kim M S, Ralph T C and Ham B S 2004 \PR A \textbf{70} 061801(R);
Paris M 1999 \textit{J. Opt. B: Quantum Semiclass. Opt.} \textbf{1}  662

\bibitem{Scripta} Adam P, Kiss T, Dar\'azs Z and Jex I 2010
\textit{Phys. Scr.} \textbf{T140} 014011

\bibitem{KerrState} Lee KS, Kim MS, Lee S-D and Bu\v zek V
1993 {\it J. Korean Phys. Soc.} \textbf{26} 197; Stobi\'nska M, Milburn GJ and W\'odkiewicz K 2008 \PR A \textbf{78} 013810



%\bibitem{Shapiro}
%Shapiro J H 2006 \PR A \textbf{73}  062305; 
%Leung P M, Ralph T C, Munro W J and Nemoto K arXiv:0810.2828v2 [quant-ph]
%
%\bibitem{inprep}Adam P, Kiss T and Jex I in preparation


%\bibitem{Janszky}  Janszky J, Vinogradov A V,  Kobayashi T and Kis Z 1994 \PR A \textbf{50} 1777 

\end{thebibliography}
\end{document}